# Globalization and long-run co-movements in the stock market for the G7: an application of VECM under structural breaks

Rui MENEZES[1,2]* & Andreia DIONÍSIO[3,4]

[1]*IUL-ISCTE Business School, Av das Forças Armadas, 1649-026 Lisboa, Portugal;*
[2]*UNIDE-IUL Research Center, Av das Froças Armadas, 1649-026 Lisboa, Portugal;*
[3]*University of Évora, Largo dos Colegiais, 2, 7000-803 Évora, Portugal;*
[4]*CEFAGE-UE Research Center, Largo dos Colegiais, 2, 7000-803 Évora, Portugal;*

This paper analyzes the process of long-run co-movements and stock market globalization on the basis of cointegration tests and vector error correction (VEC) models. The cointegration tests used here allow for structural breaks to be explicitly modeled and breakpoints to be computed on a relative-time basis. The data used in our empirical analysis were drawn from Datastream and comprise the natural logarithms of relative stock market indexes since 1973 for the G7 countries. The main results point to the conclusion that significant causal cointegration effects occur in this context and that there is a long-run equilibrium relationship that governs the worldwide process of market integration. Globalization, however, is a complex adjustment process and in many cases there is only evidence of weak market integration which means that non-proportional price transmission occurs in the market along with proportional changes. The worldwide markets, as expected, appear to be driven in general by the US stock market.

**Globalization, long-run co-movements, market integration, VECM, cointegration, structural breaks**

Recent debates on long-run co-movements and economic globalization have led to extensive research that try to determine its causes and explain the consequences of these phenomena in terms of market performance and their ability to adjust globally to economic boosts and crises. This has been relevant in the case of financial markets and in particular in stock market studies [1-9]. However, many of these studies lack a theoretical background of what is globalization and how it can be measured, and rely solely on conclusions based on correlation relationships among stock returns.

Although returns are considered the most important factor affecting investor decisions, prices may also play a role in the process of market adjustment and, in particular, in the process of market integration. In fact, while traditionally returns are considered as a complete and scale-free summary of investment opportunities and have more attractive statistical properties (*e.g.*, stationarity and ergodicity) than prices, the latter incorporate important information about the long-run characteristics of the market that are, by construction, lost in the former. Indeed, continuous compounding returns are typically computed as the log ratio between prices at dates $t$ and $t-1$ or, equivalently, $r_t = \Delta(\log P_t)$ and by taking the first difference of an observed variable $P_t$ one

*Corresponding author (email: rui.menezes@iscte.pt)

removes the long-run information contained in the data.

Besides, the current econometric technology allows us to deal with the problem of nonstationarity of prices over time in a fairly trivial way by using the concept of cointegration. And, even the problem of scaling can be easily overcome by using relative prices rather than the originally observed stock prices or indexes without any loss of generality.

Under these circumstances, we argue that in order to keep the model as general and flexible as possible, researchers should keep as much original information as possible to avoid data manipulation distortions, and to allow for long- and short-run effects to work together in order to make meaningful predictions. Why should one discard the long-run information contained in the data when its econometric treatment is nowadays standard methodology?

Without the long-run information, any study on globalization and market integration based on price movements becomes virtually impossible to perform. The idea is that long-run movements of log prices highlight the equilibrium relationships between the data, whereas short-run movements of returns or log price changes capture deviations to the equilibrium relationships that also adjust in the short-run. It is possible that two data series move together in the short-run while they diverge in their long-run movements, a situation where price data do not cointegrate despite returns or price changes display significant relationships.

Long-run and short-run price movements can be modeled by using an error correction framework. One advantage of the error correction model is that it allows for historical prices and returns to affect simultaneously but separately the behavior of current stock market prices over time, under the condition of cointegration [10-12]. On this basis, one can construct statistical tests to verify whether markets are integrated and what is the direction of causality or price transmission.

In this paper VECM cointegration tests as well as exogeneity and long-run adjustment tests under structural breaks will be employed in order to investigate whether the stock markets of the G7 countries are, in some way, related in the long-run and react in a systematic way to shocks occurring in the global market. A concise discussion of the theoretical background is presented in Section 1. Section 2 discusses the methodological aspects related to the present research work. In Section 3 we present the data set used in our empirical analysis and the main results that were obtained. Finally, Section 4 presents a summary discussion of the main conclusions.

## 1 Theoretical background

Globalization, in its literal sense, is the process of transformation of local or regional phenomena into global ones and can be described as a process by which the world population is gradually more integrated into one sole society. That is, globalization implies uniformity in terms of tastes, behaviors, prices, goods accessibility, and much more. It is a process of interaction among the economic and social agents (people, firms, etc.) driven by international trade and investment and aided by information technology that reduced significantly the geographical distance barriers and communication difficulties between people living in different parts of the world.

One important aspect of economic globalization is market integration. In the sense of Stigler [13] and Sutton [14] a market is "the area within which the price of an asset tends to uniformity after allowing for different transportation costs, differences in quality, marketing, etc". This definition relates the price evolution in the long-run, although deviations may occur in the short-run. It is, therefore, an equilibrium relationship or long-run trend.

On the other hand, market integration refers to proportionality of price movements over time for an asset or group of assets. The economic variable price is, therefore, a key element in the process of market globalization and provides a suitable framework for testing market integration by looking at the price relationship of assets over time. Strictly speaking we should look at proportionality of price movements over time for a given asset sold in geographically separated markets in order to show whether these markets are integrated or not. This is what we may call *strong market integration* but, in many cases, market integration only occurs in a weak or imperfect way. If this is so, one can expect nonlinearities and other types of price distortions to be present in the process of price transmission and a test of *weak market integration* can be performed on the basis of causality between prices, independently of whether they are proportional or not over time.

For example, a shock in the US market, usually considered as the dominant driver market, may be transmitted in quite different manners to the remaining markets, in which case it is difficult to conclude that markets tend to uniformity. This is not compatible with strong market integration but fits very well in the notion of weak market integration. Indeed, the process of market globalization is complex and the nonlinear transmission of price movements must be properly accommodated within the context of stock market globalization [15-18].

The definition of strong market integration presented above implies that the Law of One Price (LOP) holds. This means that there is not only a causal relationship between prices but they must also be proportional over time. This law, described by Cassel [19] and other prominent economists such as Cournot and Marshall, can be regarded as a special case of the following ADL($p$, $q$) price relationship:

$$x_{1t} = \theta + \sum_{k=1}^{p} \rho_k x_{1,t-k} + \sum_{j=0}^{q} \beta_j x_{2,t-j} + v_t, \quad (1)$$

where $x_{it}$ ($i$ = 1, 2) denotes the relative prices (measured in logs) of asset $i$ at time $t$, $\rho_k$ captures the extent of autocorrelation in $x_{1t}$, $\beta_j$ measures the relationship between prices (in levels and lags) and $v_t$ is a white noise perturbation.

One can say that $x_{2t}$ causes $x_{1t}$ if H$_0$: $\beta_j = 0$, $\forall j = 0, \ldots, q$ is rejected; notice that the relationship can be bidirectional. If there is just one unidirectional causal relationship, then one of the markets can effectively influence the other market prices but the reverse is not true. If the null hypothesis is not rejected in both cases, then there is no causal relationship between the underlying prices and one can say that they do not belong to the same market space. A generalization of this relationship to more than two price variables is fairly trivial.

A test of the LOP based on (1) can be set up from the restriction:

$$\sum_{k=1}^{p} \rho_k + \sum_{j=0}^{q} \beta_j = 1, \quad (2)$$

where, under the null, one says that there is strong long-run market integration. If $\beta_0 = 1 \wedge \beta_j = \rho_k = 0$, $\forall k, j > 0$, there occurs strong instantaneous market integration and expression (1) reduces to a static or contemporaneous (also called long-run equilibrium) model with no lagged effects.

If the null hypotheses of no causal relationship and strong market integration are rejected, the variables are nonlinearly related and the sum of $\rho_k$ and $\beta_j$ captures the degree of nonlinearity between them. It is important to note that (1) comes out from a multiplicative model of the original (no logarithmic) price variables and, under the estimated solution of the model, the price function is homogeneous where $m^{\Sigma \rho + \Sigma \beta}$ is the multiplicative factor of $\exp(x_{1t})$ when $\exp(x_{2t})$ is multiplied by $m$.

The $\exp(\theta)$ parameter measures the intertemporal relationship between prices and is, in our case, a constant. It has a direct economic interpretation when there is market integration and the LOP holds. In the case of the estimated static model, if $\theta = 0$, prices are alike. If $\theta \neq 0$, prices move proportionally but may differ due to different transportation costs, quality, etc. The first case is known as the strong version of the LOP and the second case is the corresponding weak version.

In finance, the LOP is frequently described in the context of the Arbitrage Pricing Theory [20] in terms of price equality independently of the means used to create the underlying asset and it can be said that the LOP and the Purchasing Power Parity (PPP) are equivalent concepts. In particular, the notion of absolute PPP is equivalent to the strong version of the LOP whereas the relative PPP is equivalent to the weak version of the LOP.

Under the PPP conditions, model (1) can be estimated using data converted to the same currency. As noticed before, if all lagged relationships are non-significant we have the following long-run equilibrium model (attractor):

$$v_t = x_{1t} - \theta - \beta_0 x_{2t}. \quad (3)$$

A test of the LOP using (3) can be carried out under the null hypothesis that $\beta = (1, -\theta, -1)$, where $\beta$ is the vector of right-hand side parameter estimates of the model normalized to $x_{1t}$ and $\beta_0$ is the elasticity of price transmission. Recall that if $\theta \neq 0$, the weak version of the LOP holds; otherwise one faces the strong version of the LOP.

## 2 Methodological issues

Estimation of (3) is straightforward if the variables in the model are stationary and there is no residual autocorrelation, since the OLS estimates converge asymptotically to the true values of the parameters. In the latter case the problem is overcome using an adequate ADL specification such as (1). However, in the case of nonstationary variables, the solution is not as simple as that. It is, therefore, important to know the stationarity properties of $x_{it}$ before proceeding to the estimation of any regression model linking them. In general, a stochastic time series is said to be strictly stationary if its joint probability distribution is invariant over time. This means that one needs to analyze all the moments of the joint probability distribution which is impossible or, at least, rather cumbersome in many cases. Alternatively, the second order properties of the distribution are a sufficient characterization of the joint probability distribution in the case of the multivariate normal distribution. Stationarity of $x_{it}$ under these conditions is known as weak (or covariance) stationarity, where:

$$\begin{aligned} E(x_{it}) &= \mu & \forall t = 1, \ldots, T \\ Cov(x_{it}, x_{i,t-k}) &= \gamma_k & \forall t = 1, \ldots, T \wedge k = 0, 1, 2, \ldots \end{aligned} \quad (4)$$

There are two opposite cases of nonstationarity: (1) deterministic nonstationarity or TSP (trend stationary process) and (2) stochastic nonstationarity or DSP (difference stationary process). Deterministic nonstationarity can be modeled by including a linear or nonlinear deterministic term in the regression equation. Modeling stochastic nonstationarity is more intricate as, in this case, stationarity can only be induced by taking the first or higher order differences of the original data, at the expense of losing the long-run information contained in the original data. In practice, however, it is quite common to find situations where a process combines both types of nonstationarity.

Stochastic nonstationarity can be detected on the basis of unit root tests, of which the most popular is the Augmented Dickey-Fuller or ADF test [21, 22] based on the following data generation process:

$$\Delta x_{it} = \mu_0 + \mu_1 t + (\rho - 1)x_{i,t-1} + \sum_{k=1}^{p} \gamma_k \Delta x_{i,t-k} + \varepsilon_t, \quad (5)$$

where $\mu_0$ is a constant term, $\mu_1 t$ is a linear deterministic trend in the data, $(\rho - 1) x_{i,t-1}$ denotes the corresponding sto-chastic trend and the errors $\varepsilon_t \sim \text{iid}(0, \sigma^2_\varepsilon)$. The symbol $\Delta$ denotes a first difference, as usual, and the summation term captures any autocorrelation of the left-hand side variable. Taking $\mu_1 = \gamma_k = 0$, the ADF equation reduces to an AR(1) process. The null hypothesis in the ADF test is $\rho = 1$, denoting stochastic nonstationarity, and the testing procedure uses the MacKinnon [23, 24] critical values.

Despite their popularity, the ADF tests suffer from low power problems when the process is stationary with roots close to one [25]. Additionally, some unit root processes behave more like a white noise than like a random walk in finite samples. For this reason, it is convenient to use alternative tests in order to conclude more accurately about the stationary nature of the series under analysis. An alternative to the ADF test is the KPSS [26] which postulates as the null hypothesis that the time series is trend stationary, against the alternative that it contains a stochastic trend.[1] The data generation process of the KPSS test is given by:

$$x_{it} = \mu t + z_t + u_t \qquad (6)$$
$$z_t = z_{t-1} + \varepsilon_t$$

where $x_{it}$ is the sum of a deterministic trend ($\mu t$), a random walk ($z_t$) and a stationary error variable ($u_t$) and where $\varepsilon_t \sim \text{iid}(0, \sigma^2_\varepsilon)$. The null hypothesis of stationarity is given by $\sigma^2_\varepsilon = 0$, where the initial value $z_0$ is a constant. Given that $u_t$ is a stationary error variable, then under the null $x_{it}$ is a TSP.

Most economic and financial time series that are stochastically nonstationary are also integrated of first order, that is, differencing once is enough to achieve stationarity. It is the case of the vast majority of stock market price series and indexes. Suppose, for instance, that one is interested in the long-run properties that rule the relationship between two or more first-order integrated price series. In this case, one needs to focus the analysis on the variables measured in levels. However, a linear combination of first-order integrated variables usually generates a residual variable that is also first-order integrated. Under these circumstances, the usual $t$ and $F$ tests carried out on the OLS estimates do not follow, respectively, the $t$ and $F$ distributions and these estimates are, thus, meaningless [27]. What the model is most possibly capturing is a common stochastic trend between the variables in levels and not a causal relationship as required. This is known in the literature as the spurious regression problem [28]. Additionally, the residuals are strongly autocorrelated and the Durbin-Watson statistic converges to zero. Thus, the time series being analyzed are not related in the long-run although they may be related in the short-run. A special case of this is the relationship between two random walks.

---

[1] Many other unit root or stationarity tests have been proposed in the literature (*e.g.*, Phillips-Perron - PP, Elliott, Rothenberg and Stock - Point Optimal, Elliott, Rothenberg and Stock – DFGLS, Ng and Perron – NP).

Albeit in general a linear combination of nonstationary variables generates residuals that are also nonstationary, there is a special case where the residuals thus obtained are stationary. This is the case when the variables in the model are said to be cointegrated [11]. In this case, the OLS estimator of $\beta_0$ in (3) is super-consistent, converging to its actual value more quickly than if the variables were stationary [29]. Under these circumstances, the usual $t$ and $F$ statistics remain valid for testing hypotheses about the parameters. One simple way to test for cointegration is to regress $x_{it}$ on $x_{jt}$ ($i \neq j$) and then analyze whether the resulting residuals are stationary. Note that if the variables are cointegrated then $\beta_0 \neq 0$ and the cointegration test can be interpreted in terms of market integration. However, which variable should be considered as endogenous and which should be considered as exogenous? Isn't it possible that they are both endogenous and thus exert a mutual influence on each other? In this case biases may occur due to endogeneity [30] and a multi-equation model would be preferable to use instead of the single equation model given in (3). Furthermore, since the dynamic terms are omitted in model (3), one would expect that the residuals $v_t$ are autocorrelated.

An alternative to the single equation models presented in (1) and (3) is based on the specification of a Vector Autoregression (VAR) or, under the hypothesis of cointegration, a Vector Error Correction Model (VECM) of the type:

$$\Delta \mathbf{x}_t = \boldsymbol{\alpha}\boldsymbol{\beta}' \mathbf{x}_{t-1} + \sum_{k=1}^{p-1} \boldsymbol{\Gamma}_k \Delta \mathbf{x}_{t-k} + \boldsymbol{\mu} + \boldsymbol{\varepsilon}_t, \qquad (7)$$
$$\boldsymbol{\alpha}\boldsymbol{\beta}' = \sum_{k=1}^{p} \mathbf{A}_k - \mathbf{I} \,;\, \boldsymbol{\Gamma}_k = -\sum_{j=k+1}^{p} \mathbf{A}_j,$$

where $\mathbf{x}_t$ is an $i$-dimension vector of nonstationary endogenous variables, given in levels, and representing, for instance, the natural logarithms of relative asset prices (*e.g.*, stock indexes), $\boldsymbol{\mu}$ is an $i$-dimension vector of constants and $\boldsymbol{\varepsilon}_t$ denotes an $i$-dimension vector of errors or perturbations where $\boldsymbol{\varepsilon}_t \sim \text{iid}(0, \boldsymbol{\Sigma})$. The $\mathbf{A}_k$ denote $p$ $i$-order matrices of parameters where each of them is associated with an $i$-dimension vector of lagged endogenous variables up to order $p$. The variance-covariance matrix $\boldsymbol{\Sigma}$ is definite positive and the errors $\boldsymbol{\varepsilon}_t$ are not serially correlated since the dynamic process linking the data is explicitly specified in the model, although they may be contemporaneously correlated. This system specification contains information about the short- and the long-run adjustment parameters through the estimates of $\boldsymbol{\Gamma}_k$ and $\boldsymbol{\alpha}\boldsymbol{\beta}'$, respectively.

The use of the VEC model in the context of cointegration is assured by the Granger Representation Theorem, which states that "if there exists a dynamic linear model with stationary perturbations and the data are first-order integrated, then the variables are cointegrated of order CI(1, 1)".

If $\mathbf{x}_t \sim I(1)$, where I(1) means nonstationary but integrated of first order, then $\Delta \mathbf{x}_t \sim I(0)$, where I(0) means stationary or integrated of order zero, and $\boldsymbol{\Gamma}_k \Delta \mathbf{x}_{t-k} \sim I(0)$. The term $\alpha\beta'\mathbf{x}_{t-1}$ is a linear combination of I(1) variables which in turn is I(0) under the assumption of cointegration, where $\alpha$ represents the speed of adjustment to equilibrium and $\beta$ is the matrix of long-run coefficients, that is, the cointegrating vectors. This is true when there are $r$ cointegrating vectors (with $0 < r < i$) representing the error correction mechanism in the VEC system. Notice that $r$ denotes the rank of $\alpha\beta'$. Once obtained the cointegrating relationships and the matrices $\alpha$ and $\beta$ are estimated, the VEC can be fully estimated incorporating those vectors.

For a set of variables to be cointegrated is necessary that there is at least one cointegrating vector in the system ($r \geq 1$). However, the matrix $\alpha\beta'$ cannot be regular since in this case all vectors contained in $\alpha\beta'$ are linearly independent. Thus, if $r = i$, $\mathbf{x}_t$ is a vector of stationary variables. If $r = 0$, then $\alpha\beta' = \mathbf{0}$ and the variables in $\mathbf{x}_t$ are not cointegrated since there is none cointegrating vector or long-run relationship linking the variables in levels. In this case, current returns or price changes are only a function of previous returns or price changes; prices play no role in the overall system and long-run predictions as well as tests of market integration based on price theory cannot be made. Finally, under cointegration, $r$ indicates how many independent long-run relationships exist in the system. In this context, historical prices and returns can be used in the model, which is preferable to using just stock returns since the former retain both the long-run and the short-run information contained in the data, while the latter only capture the short-run information.

As noticed, the VEC model represented in (7) can be interpreted as a relationship between prices and returns in a given market. What it says is that current returns or price changes are a linear function of previous returns or price changes and historical prices. Such historical prices form a long-run or equilibrium relationship where the involved variables co-move over time independently of the existence of stochastic trends in each of them, so that their difference is stable. The long-run residuals measure the distance of the system to equilibrium at each moment $t$, which may be due to the impossibility of the economic agents to adjust instantaneously to new information or to the short-run dynamics also present in the data. There is, therefore, a whole complex adjustment process involving short-run and long-run dynamics when the variables are cointegrated.

## 3  Data and results

The data set used in our empirical analysis consists of daily stock price series representing the G7 countries: US, Canada, Japan, UK, Germany, France and Italy. These markets represent the bulk of transactions in worldwide stock markets. The data are the relative price indexes for these markets and were collected from the Datastream database covering the period from January, 1st 1973 (base 100) to January, 21st 2009, totalizing 9408 daily observations (five days per week) for each series (total of 65856 data points). Figure 1 shows a plot of the seven series in relative prices over the period analyzed.

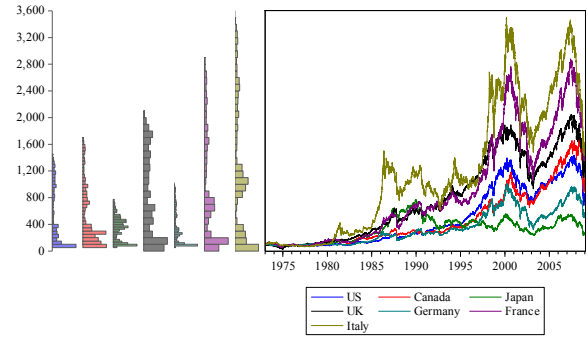

**Figure 1**  Relative stock price indexes for the G7 countries (Datastream).

A well-known problem with worldwide daily data is that trading days vary across markets, as they operate in different time zones. Thus, in order to correct this synchronization bias we use the procedure proposed by Beine et al. [31]. Since the Japanese market is the first to close amongst the G7 stock markets we use a one-day lagged effect from the other markets towards the Japanese one to capture contemporaneous relationships. On the other hand, since the North-American markets are the last to close, a one-day lagged effect from the North-American stock markets towards the other markets will capture any contemporaneous relationship among them.

It is remarkable how similar the time-path pattern looks for these seven stock market indexes with market boosts and crises apparently synchronized for all the countries. However, data dispersion increases substantially along time, especially after the oil and energy crises of the seventies and, further on, since the end of the 20th century. Price volatility over the period was substantially higher for Italy, France and the UK than for Canada, the US, Germany and Japan.

Notice that all series are flatter than the Gaussian distribution and right-skewed (see the histograms on the left hand-side of Fig. 1), therefore the null hypothesis of normality is rejected for all of them. This is typical of stock market price series as well as leptokurtosis and fat tails is usually observed in returns data. From this point onwards the analysis will only consider the natural logarithms data, that is, stock prices actually refer to the natural logarithms of the relative price indexes and stock returns or price changes denote the difference between log relative prices at two adjacent dates.

Before proceeding to the analysis of market integration one should look at the (non)stationary nature of the G7 se-ries. Unit root and stationarity tests in levels and in first differences for all the series are shown in Table 1:

**Table 1** Unit root and stationarity tests in levels and in first differences

| Variable | ADF [a, c, d] | KPSS [b, c, d] |
|---|---|---|
| US[f] | -1.709328 | 0.960241 ** |
| Canada[e] | -2.806501 | 0.468607 ** |
| Japan[f] | -0.269712 | 2.549435 ** |
| UK[g] | -0.736909 | 2.320246 ** |
| Germany[e] | -1.722877 | 0.568883 ** |
| France[e] | -1.050611 | 1.038102 ** |
| Italy[g] | -0.500341 | 1.661498 ** |
| $\Delta$US | -70.39091 ** | 0.244395 |
| $\Delta$Canada | -88.91458 ** | 0.075838 |
| $\Delta$Japan | -69.26301 ** | 0.126957 |
| $\Delta$UK | -45.20940 ** | 0.220531 |
| $\Delta$Germany | -92.36380 ** | 0.130575 |
| $\Delta$France | -89.32861 ** | 0.186063 |
| $\Delta$Italy | -44.66293 ** | 0.302849 |

Notes: [a] MacKinnon (1996) critical values: -3.43 (1%) and -2.86 (5%) for constant and -3.96 (1%) and -3.41 (5%) for constant and linear trend. [b] Kwiatkowski-Phillips-Schmidt-Shin (1992) critical values: 0.739 (1%) and 0.463 (5%) for constant and 0.216 (1%) and 0.146 (5%) for constant and linear trend. [c] exogenous terms in levels: constant and linear trend. [d] exogenous terms in 1st differences: constant (except for Japan in the KPSS test which is constant and linear trend). [e] 1 lag in levels for ADF. [f] 2 lags in levels for ADF. [g] 4 lags in levels for ADF. ** significant at 1%.

As noted before, the ADF and KPSS tests are designed to capture weak stationarity with opposite null hypotheses. In the former case the null hypothesis of nonstationarity of the variables in levels is not rejected but it is rejected at 1% for the variables in first differences. In the latter case the null hypothesis of stationarity in levels is rejected at 1% but it is not rejected in first differences. The results are, therefore, consistent in both cases and lead to the conclusion that the price series under analysis are, in fact, integrated of first order. The number of lags selected in each test was set on the basis of the SBC information criterion [32]. One can thus conclude that the stock price series under analysis are nonstationary variables while stock returns are stationary. Therefore, the analysis of stock market integration has to be conducted within the environment of cointegration.

The typical approach to cointegration in the context of multi-direction endogenous relationships is the specification of a VEC model, as described before. Testing procedures usually rely on the Johansen methodology which provides more powerful test statistics than the original EG methodology in the presence of endogenous systems. The Johansen tests are based on the rank of matrix $\alpha\beta'$, but also suffer from low power problems in the presence of structural breaks because they assume that the cointegrating vector is time-invariant under the alternative hypothesis. This may lead to an over non-rejection of the null that the rank of $\alpha\beta'$ is zero, thus leading the researcher to incorrectly conclude that there is no long-run relationship in the data. Structural breaks are likely to occur especially when we analyze relatively long time series, as is our case (daily data over 36 years), with shifts or the occurrence of extreme events (one or several) somewhere within the period. Thus, formal tests for structural breaks are required.

To examine the statistical presence of structural breaks, we first performed CUSUM and CUSUM-Q tests. As usual, we tested for the stability of our time series by regressing each of them on a nonsignificant constant. The results indicate the presence of structural breaks for all the variables and some of these structural breaks seem to be related to the occurrence of stock market crashes and financial crises, reinforcing the possibility of nonlinear behavior among the stock markets under study. These tests, however, are best suited under a stationary environment.

Under nonstationary structural breaks, an alternative to the Johansen method must be devised in order to test for cointegration without loss of statistical power. To this end, the Phillips and ADF tests proposed by Gregory and Hansen [33] are likely to constitute a good approach. The GH test statistics for cointegration are presented in Table 2:

**Table 2** ADF and Philips cointegration tests

| Variable | Statistic | Breakpoint |
|---|---|---|
| **ADF*** | | |
| C | -6.066 ** | (0.3911) |
| C/T | -6.282 ** | (0.3911) |
| C/S | -7.020 ** | (0.3472) |
| **$Z_t$*** | | |
| C | -6.155 ** | (0.3912) |
| C/T | -6.379 ** | (0.3912) |
| C/S | -7.061 ** | (0.3473) |
| **$Z_\alpha$*** | | |
| C | -77.154 ** | (0.3912) |
| C/T | -82.315 ** | (0.3912) |
| C/S | -96.400 ** | (0.3091) |

Notes: [a] Gregory and Hansen (1996) critical values: -6.05 (1%) for C, -6.36 (1%) for C/T and -6.92 (1%) for C/S for the ADF* and $Z_t$* tests; -70.18 (1%) for C, -76.95 (1%) for C/T and -90.35 (1%) for C/S for the $Z_\alpha$* test. [b] $m = 6$. ** significant at 1%.

The GH approach relies on a regression equation of the form:

$$x_{1t} = \mu_0 + \mu_1 \varphi_{t\tau} + \beta t + \alpha_0' x_{2t} + \alpha_1' x_{2t} \varphi_{t\tau} + u_t, \qquad (8)$$

where the observed data is $x_t = (x_{1t}, x_{2t})$, $x_{2t}$ is an $m$-vector of variables, $u_t$ is a stationary error term, $\beta$, $\mu_s$, $\alpha_s$ ($s = 0, 1$) are parameters or vectors of parameters, $t = 1, \ldots, n$, and $\varphi_{t\tau}$ is a dummy variable that governs the structural change, so that:

$$\varphi_{t\tau} = \begin{cases} 0 & \text{if} \quad t \leq [n\tau] \\ 1 & \text{if} \quad t > [n\tau] \end{cases}, \qquad (9)$$

where $\tau \in (0, 1)$ and [ ] denotes integer part. The unknown parameter $\tau$ symbolizes the relative timing of the break point [33].

If $\beta = 0$ and $\alpha_1 = 0$, the model is called a level shift (C) denoting just a change in the intercept $\mu$. A level shift with trend occurs when $\alpha_1 = 0$ and $\beta \neq 0$ (C/T). Finally, when $\beta = 0$ and $\alpha_1 \neq 0$ the model is called a regime shift (C/S), allowing the slope vector to shift as well.

The results reject the null hypothesis of no cointegration at the 1% level in all the tests and indicate that there is a breakpoint in the series at about 39% of the time period analyzed for the level shift and level shift with trend. A regime shift also occurs at about 31%-35% of the period. The level shift occurred in 1987 (black Monday) and the regime-shift occurred in 1984-85. There is, therefore, notable evidence that stock markets among the G7 are linked in the long-run, despite the incidence of shifts provoked by structural breaks resultant from crashes or other economic or financial distortions that change the market behavior in a "permanent" way. A first, but not surprising, conclusion is that stock market integration does happen for the G7, although at this point one cannot say whether this occurs in a weak or strong form.

On the light of the above cointegration tests and under the Granger Representation Theorem it is possible to specify a VEC model with the guarantee that the residuals are stationary as required for inference purposes. Under these circumstances, the parameter estimates can be computed as usual and the $t$ and $F$ statistics can be used for testing purposes in the same way as in the case of a stationary OLS or ML model. The VEC model allows for testing parameter restrictions on the $\alpha\beta'$-matrix that contains the long-run system information.

On the basis of the VEC estimates it is possible to test for the direction of causality that occurs in the system [10, 34]. A popular test used in this context is the Granger causality [35]. Simple manipulation of the VEC leads to a reparameterized version of (7) where the vector $\mu$ is multiplied by the estimated long-run residuals and the matrices $\mathbf{A}_i$ ($i = 1, \ldots, m$) contain the coefficients of the lagged returns for each variable separately. For a two cointegrated variable system and $p$ lags,[2] and noting that $\hat{u}_{t-1} = \hat{\boldsymbol{\beta}}'\mathbf{x}_{t-1}$ one has:

$$\Delta \mathbf{x}_t = \mathbf{A}_1 \Delta \mathbf{x}_{1,t-j} + \mathbf{A}_2 \Delta \mathbf{x}_{2,t-j} + \boldsymbol{\mu}\hat{u}_{t-1} + \boldsymbol{\varepsilon}_t, \qquad (10)$$

where $\Delta \mathbf{x}_t$ represents returns or log price changes at time $t$ and $\Delta \mathbf{x}_{i,t-j}$ ($i = 1, 2; j = 1, \ldots, p-1$) denotes lagged returns up to $p-1$ of the $i^{th}$ variable. $\mathbf{A}_1$ and $\mathbf{A}_2$ are $[2\times(p-1)]$ matrices. $\boldsymbol{\mu}$ and $\boldsymbol{\varepsilon}_t$ are (2×1) vectors and $\hat{u}_{t-1}$ denotes the long-run residuals, where $u_t \sim I(0)$. A Granger causality test can be carried out on the basis of the null hypothesis: $\delta_{i1} = \ldots = \delta_{i,p-1} = \mu_i = 0$, where the $\delta_i$ coefficients correspond to the $i^{th}$ row of $\mathbf{A}_2$. The test then compares the mean squared error

---

[2] Notice, however, that the number of lags can be different for each variable.

under the null and under the alternative hypotheses, such that:

$$MSE(\hat{x}_{1t} \mid I_{t-1}) < MSE(\hat{x}_{1t} \mid I_{t-1} \setminus Ix_{2,t-1}), \qquad (11)$$

where MSE is the mean squared error, $I_{t-1}$ represents the set of all past and present information existing at date $t-1$, $Ix_{2,t-1}$ represents the set of all past and present information existing on $x_2$ at date $t-1$, i.e., $Ix_{2,t-1} = \{x_{21}, x_{22}, \ldots, x_{2,t-1}\}$, $x_{1t}$ is the value of $x_1$ at moment $t$ ($x_{1t} \subset I_t$) and $\hat{x}_{1t}$ is a non biased predictor of $x_{1t}$. The Granger causality test is interpreted as follows: $x_{2t}$ Granger causes $x_{1t}$ if, *ceteris paribus*, the past values of $x_{2t}$ help to improve the current forecast of $x_{1t}$. On the other hand, $x_{2t}$ instantaneously causes $x_{1t}$ in the sense of Granger if, *ceteris paribus*, the past and present values of $x_{2t}$ help to improve the prediction of the current value of $x_{1t}$, that is:

$$MSE(\hat{x}_{1t} \mid I_t \setminus x_{1t}) < MSE(\hat{x}_{1t} \mid I_t \setminus Ix_{2,t}, x_{1t}). \qquad (12)$$

In practice, the Granger causality test performed in statistical software postulates as the null hypothesis that "$x_{2t}$ does not Granger cause $x_{1t}$", so that causality implies the rejection of the null.

Table 3 presents the Granger causality tests for the variables in levels, that is, stock prices. Here, $x_{2t}$ represents the variables in the first column and $x_{1t}$ represents the variables in the first row. One can say, therefore, that for the significant causal relationships the historical prices of the former market affect the current price of the latter, forming a dynamical long-run relationship in the global economy. As we can see, about 74% of the coefficients are statistically significant, which means that there is substantial long-run causal effects among these markets, of which many of them are feedback relationships. However, we found no causal relationship in any direction for the pairs Germany-France and Germany-Italy which is a surprising result since we expected that European stock markets would co-move quite closely given the environment and political rules of the European Union, formerly European Community.

Another important result is that, in the long-run, the US causes more than is caused by other markets. To see this, note that the $F$-statistics of the former (1$^{st}$ row) are substantially larger than the $F$-statistics of the latter (1$^{st}$ column). This is consistent with the idea that the US stock market, to a greater extent, 'exports' more than 'imports' boosts and crises, being therefore the engine of the global financial world. For example, a crisis with origin in the US can spread in a broader way to other markets (as it seems in the recent crisis) than a crisis with origin in Japan or even any European country. Canada shows an overall picture very similar to the US, that is, in general it causes more other markets than is caused by them, except in what refers to the US. Canada, however, appears to be caused only by the US, Japan and, to a lesser extent, the UK. Conversely, Japan is the most endogenous of the G7 markets. The European countries do not show an overall systematic pattern of causality, though the UK appears to emerge like an attractor in the EU context (but not with France) and follows the North-American markets. This is also surprising insofar we would expect Germany to be the leading European stock market, given its role as the 'head' of the European Union economy, albeit one should recognize the very important role of the London Stock Exchange in the global financial world.

**Table 3**　Granger causality for log prices

| Variable | US | | Canada | | Japan | |
|---|---|---|---|---|---|---|
| US | - | | 142.898 | ** | 716.963 | ** |
| Canada | 36.0065 | ** | - | | 361.477 | ** |
| Japan | 14.3828 | ** | 4.86702 | ** | - | |
| UK | 8.91317 | ** | 3.99597 | * | 233.251 | ** |
| Germany | 6.03121 | ** | 2.29723 | | 284.715 | ** |
| France | 9.72877 | ** | 1.51540 | | 259.915 | ** |
| Italy | 1.93860 | | 1.01208 | | 107.911 | ** |

| Variable | UK | | Germany | | France | |
|---|---|---|---|---|---|---|
| US | 475.981 | ** | 390.273 | ** | 470.843 | ** |
| Canada | 113.420 | ** | 73.5093 | ** | 117.433 | ** |
| Japan | 27.3334 | ** | 21.9847 | ** | 24.1686 | ** |
| UK | - | | 7.33226 | ** | 6.02136 | ** |
| Germany | 1.13299 | | - | | 2.17821 | |
| France | 6.42979 | ** | 1.54816 | | - | |
| Italy | 0.61978 | | 1.42051 | | 7.52491 | ** |

| Variable | Italy | |
|---|---|---|
| US | 156.006 | ** |
| Canada | 46.4521 | ** |
| Japan | 7.23094 | ** |
| UK | 3.54475 | * |
| Germany | 0.32732 | |
| France | 1.21910 | |
| Italy | - | |

Notes: H$_0$: $x_{it}$ does not Granger cause $x_{jt}$ ($i \neq j$). 2 *lags*. 9406 observations in each series. ** significant at 1%. * significant at 5%.

One may now proceed to the analysis of long-run market integration between market prices and returns in order to investigate whether the LOP holds for the system as a whole and/or for some of these markets bilaterally. Notice that if the LOP holds, strong market integration occurs in the system; otherwise there is weak market integration given that the variables are cointegrated, as shown before. The testing procedure considers the VAR estimates of the seven-country system in order to investigate whether price transmission is proportional or not. This test utilizes the sum of the VAR estimates for each equation in the system, where the null is H$_0$: $\sum_{h=1}^{ip} a_{ih} = 1$, ($a_{ih} \in \mathbf{A}_k$, $k = 1,\ldots,p$). The results are presented in Table 4.

**Table 4**　Long-run market integration tests for multivariate systems

| Variable | $\sum a_{ih}$ | β | Std. error | |
|---|---|---|---|---|
| US | 0.99898 | 1.00000 | | |
| Canada | 0.99962 | 0.23742 | 0.11652 | * |
| Japan | 0.99858 | 0.59418 | 0.08353 | ** |
| UK | 1.00046 | -0.44914 | 0.10882 | ** |
| Germany | 1.00072 | -0.06563 | 0.13967 | |
| France | 1.00149 | -0.83523 | 0.15673 | ** |
| Italy | 0.99960 | -0.03897 | 0.07929 | |

Notes: H$_0$: $\sum a_{ih} = 1$. Exogenous terms in CE: constant. 2 *lags* in the endogenous variables. 9405 observations after adjustments. ** significant at 1%. * significant at 5%.

As can be seen, the sum of the VAR estimates is very close to one in all cases, suggesting (apparently) that strong long-run market integration occurs for the G7 system. These results may be checked using the long-run VEC estimates. The null hypothesis in this test postulates that $\beta_{11} = 1$ and $\sum_{k=2}^{7} \beta_{1k} = -1$. Under the null hypothesis, the test statistic is $\chi^2(1) = 16.77231$ ($p = 0.000042$) and therefore the null is rejected (the estimated β coefficients are reported in Table 4 along with the corresponding standard errors). This implies that market integration for the G7 only occurs in a weak form, that is, although some price transmission relationships may actually be linear (proportional transmission), there are cases where they are nonlinear (non-proportional transmission) turning stock market interactions a complex system.

While being important to determine whether market integration occurs for the system as a whole, it is also important to know if shocks to the system propagate uniformly over the seven markets in analysis, in order to understand to what extent a shock affects the performance of the stock market. To this end, bivariate market integration tests must be conducted. The results of these tests are displayed in Table 5.

The test statistic follows a $\chi^2$ distribution with one degree of freedom. The null hypothesis in this test indicates proportionality or full price transmission over time. If the null hypothesis is not rejected then market *j* and market *k* are strongly integrated in the long-run. The null hypothesis is rejected in 62% of the bivariate relationships considered. Proportionality was only found for US-France, Canada-UK, Canada-Germany, Canada-Italy, Japan-UK, Japan-France, UK-Germany and Germany-Italy. For the remaining cases it was found that, in terms of pairwise relationships, although the underlying markets belong to the same market space, they are not related in a linear way. Such relationships are, therefore, much more complex and certainly some type of nonlinear relationship links the markets in the global world. For example, a shock in the US market is transmitted in a linear way only to France, while the transmission to other markets may be amplified or attenuated according to the impact factors given by the long-run coefficients.

**Table 5** Long-run market integration tests for bivariate systems

| Variable 1 ($\beta_{1j}$) | Variable 2 ($\beta_{1k}$) | $\chi^2(1)$ | $p$-value |
|---|---|---|---|
| US | Canada | 26.35202 | 0.00000 ** |
| US | Japan | 25.89823 | 0.00000 ** |
| US | UK | 9.07521 | 0.00259 ** |
| US | Germany | 20.90138 | 0.00001 ** |
| US | France | 0.45477 | 0.50008 |
| US | Italy | 14.36033 | 0.00015 ** |
| Canada | Japan | 16.68239 | 0.00004 ** |
| Canada | UK | 1.12993 | 0.28779 |
| Canada | Germany | 0.48726 | 0.48515 |
| Canada | France | 8.14178 | 0.00433 ** |
| Canada | Italy | 1.79485 | 0.18034 |
| Japan | UK | 1.27216 | 0.25936 |
| Japan | Germany | 5.64007 | 0.01756 * |
| Japan | France | 1.35996 | 0.24354 |
| Japan | Italy | 18.71836 | 0.00002 ** |
| UK | Germany | 3.79843 | 0.05130 |
| UK | France | 23.26063 | 0.00000 ** |
| UK | Italy | 6.86539 | 0.00879 ** |
| Germany | France | 13.69186 | 0.00022 ** |
| Germany | Italy | 0.23696 | 0.62641 |
| France | Italy | 19.50850 | 0.00001 ** |

Notes: H$_0$: $\beta_{1j} = 1$, $\beta_{1k} = -1$ ($j \neq k$). Exogenous terms in CE: constant. 2 *lags* in the endogenous variables. 9405 observations after adjustments. ** significant at 1%. * significant at 5%.

## 4 Discussion and Final remarks

This paper analyzes stock market globalization on the basis of market integration among the G7 countries by using a time-series approach based on logarithmic price relationships. Strong market integration occurs if shocks in one market are linearly (or proportionally) transmitted to another market (or other markets). If transmission is nonlinear (or non-proportional) one can say that weak market integration holds. These definitions, of course, only apply if there is a non-spurious relationship between the variables under study. Under nonstationarity, a cointegration system framework is used in order to obtain the relevant statistics. Structural breaks are also explicitly modeled. The results point to the following conclusions:

1. There is an overall non-spurious relationship between the seven markets analyzed, given the rejection of the null of no cointegration. These markets thus belong to the same market space.
2. The US market leads the G7 market space and the UK emerges as a regional attractor within the European context, which, in turn, is strongly affected by the North-American markets.
3. Canada may benefit from its proximity to the US where, surely, intense economic relationships, some similar economic policies and firm's relationships turn up the North American countries as a unified financial block. Canada, therefore, appears to show a dominant position in the market relatively to the non-American G7 countries.
4. Japan does not emerge as a leading market within the G7 countries but this is probably due to the long-lasting economic crisis that Japan faced. Over the period analyzed Japan appears as the most endogenous market among the G7, which means that, while not affecting much the behavior of the other markets, the Japanese stock market is greatly influenced by shocks occurring elsewhere within the G7.
5. The continental European countries do not show an overall systematic pattern of causality, arising as 'secondary' players in the global arena of stock markets within the G7.
6. For the G7 system as a whole the null hypothesis of proportional price transmission is rejected. Furthermore, for more than 60% of the bivariate tests of proportionality the null is also rejected, which suggests that market integration occurs for the G7 countries in a weak form. This is important insofar shocks occurring in the leading market, the US, are transmitted in quite different manners to the other markets, which turns quite difficult to devise common policies in order to minimize the transmission of shock effects and stabilize markets overall.

As a final remark we argue that weak market integration is nothing but just an imperfect form of harmonization in the global world and that further deepening and development is needed in order to achieve strong market integration.

*The authors thank the financial support provided by FCT–Fundação para a Ciência e Tecnologia, under the grants PTDC/GES/73418/2006 and PTDC/GES/70529/2006, and also the grant FCOMP-01-0124-FEDER-007350.*